\documentclass[acmsmall,nonacm]{acmart}
\AtBeginDocument{%
  }

\newif\ifdraft
\draftfalse

\newif\ifblind

\blindtrue
\newcommand{\ie}{i.e., }
\newcommand{\eg}{e.g., }
\newcommand{\cf}{cf. }
\makeatletter
\newcommand{\etal}{~et~al\@ifnextchar.{}{.\@}}
\newcommand{\etc}{etc\@ifnextchar.{}{.\@}}
\makeatother

\ifdraft
\newcommand{\note}[1]{\textbf{\color{red}{\{#1\}}}}
\newcommand{\todo}[1]{\textbf{\color{red}{\{TODO: #1\}}}}

\newcommand{\ke}[1]{{\color{blue}{\{K: #1\}}}}
\newcommand{\na}[1]{{\color{green}{\{N: #1\}}}}

\else

\newcommand{\note}[1]{}
\newcommand{\todo}[1]{}

\newcommand{\ke}[1]{}
\newcommand{\na}[1]{}

\fi

\usepackage{prettyref}
\newrefformat{chap}{Chapter~\ref{#1}}
\newrefformat{sec}{Sec.\,\ref{#1}}
\newrefformat{sub}{Sec.\,\ref{#1}}
\newrefformat{subsec}{Sec.\,\ref{#1}}
\newrefformat{ssub}{Sec.\,\ref{#1}}
\newrefformat{eq}{Eq.~(\ref{#1})}
\newrefformat{fig}{Fig.~\ref{#1}}
\newrefformat{plot}{Fig.~\ref{#1}}
\newrefformat{ex}{Example~\ref{#1}}
\newrefformat{list}{List~\ref{#1}}
\newrefformat{tab}{Table~\ref{#1}}
\newrefformat{enum}{Case~(\ref{#1}.)}
\newrefformat{def}{Definition~\ref{#1}}
\newcommand{\pref}[1]{\prettyref{#1}}

\DeclareRobustCommand{\ISymbol}{%
  \tikz[baseline=-0.6ex,scale=0.9]{%
    \def\r{2.5mm}  % circle radius

    \node[draw,circle,fill=lightgray,minimum size=2*\r,inner sep=0pt] (A) at (0,0) {I};
    \node[draw,circle,minimum size=2*\r,inner sep=0pt] (B) at (12mm,0) {II};
    \node[draw,circle,minimum size=2*\r,inner sep=0pt] (C) at (24mm,0) {III};

    \draw[->,line width=0.4pt] (A) -- (B);
    \draw[->,line width=0.4pt] (B) -- (C);
  }%
}

\DeclareRobustCommand{\IISymbol}{%
  \tikz[baseline=-0.6ex,scale=0.9]{%
    \def\r{2.5mm}  % circle radius

    \node[draw,circle,minimum size=2*\r,inner sep=0pt] (A) at (0,0) {I};
    \node[draw,circle,fill=lightgray,minimum size=2*\r,inner sep=0pt] (B) at (12mm,0) {II};
    \node[draw,circle,minimum size=2*\r,inner sep=0pt] (C) at (24mm,0) {III};

    \draw[->,line width=0.4pt] (A) -- (B);
    \draw[->,line width=0.4pt] (B) -- (C);
  }%
}

\DeclareRobustCommand{\IIISymbol}{%
  \tikz[baseline=-0.6ex,scale=0.9]{%
    \def\r{2.5mm}  % circle radius

    \node[draw,circle,minimum size=2*\r,inner sep=0pt] (A) at (0,0) {I};
    \node[draw,circle,minimum size=2*\r,inner sep=0pt] (B) at (12mm,0) {II};
    \node[draw,circle,fill=lightgray,minimum size=2*\r,inner sep=0pt] (C) at (24mm,0) {III};

    \draw[->,line width=0.4pt] (A) -- (B);
    \draw[->,line width=0.4pt] (B) -- (C);
  }%
}

\usepackage[inline]{enumitem}
\SetEnumitemKey{:roman}{label=(\roman*)}
\SetEnumitemKey{:alpha}{label=(\alph*)}
\SetEnumitemKey{:arabic}{label=(\arabic*)}
\SetEnumitemKey{:arabicbold}{label=\textbf{(\arabic*)}}
\SetEnumitemKey{:embeddedtriangle}{label={$\blacktriangleright$},noitemsep,topsep=-1\parskip}

\usepackage{circledsteps}

\usepackage{xcolor}
\definecolor{lightblue}{HTML}{CDE3F0}
\definecolor{darkerblue}{HTML}{18789C}

\usepackage{lipsum}
\usepackage{caption}
\usepackage{subcaption}

\usepackage{listings}

\usepackage{tikz}
\usetikzlibrary{patterns}
\usetikzlibrary {shapes.geometric} 
\usetikzlibrary{arrows}
\usetikzlibrary{calc}

\DeclareRobustCommand{\circlelabel}[2]{%
  \def\r{2.5mm}
  \tikz[remember picture,baseline=(n.base)]{
    \node[circle,draw,minimum size=2*\r,inner sep=0pt] (n) {#1};
    \node[overlay] (#2) at (n.north) {}; % make a named node, not just a coordinate
    \node[overlay] (s#2) at (n.south) {}; % make a named node, not just a coordinate
  }%
}

\usepackage{acro}
\makeatletter
\@ifpackagelater{acro}{2017/02/17}{\@ifpackagelater{acro}{2019/09/23}{}{
    \let\IDeclareAcronym\DeclareAcronym
    \renewcommand{\DeclareAcronym}[2]{%
        \IDeclareAcronym{#1}{%
        #2,foreign-plural={}
        }
    }
}}{}
\usepackage{siunitx}
\usepackage{xfp}
\DeclareSIUnit{\noop}{\kern 0pt}
\newcount\tmpnum

\def\storedataA#1{\advance\tmpnum by1
    \ifx\end#1\else
    \expandafter\def\csname data\tmp\the\tmpnum\endcsname{#1}%
    \expandafter\storedataA\fi
}

\def\assert#1{\ifthenelse{#1}{}{\errmessage{ASSERT FAIL}}}
\def\getarrdata[#1]#2{\ifcsname data#2#1\endcsname\csname data#2#1\endcsname\else\errmessage{UNSET #2#1}\fi}

\def\getarr[#1]#2{\getarrdata[#1]{#2}}

\def\roundprefixprefix[#1]#2{\SI[scientific-notation = engineering, exponent-to-prefix = true, round-mode=places,round-precision=#1]{#2}{\noop}}

\def\roundprefix[#1]#2{\ifthenelse {#2 < 1000}{#2\,\,~}{\roundprefixprefix[#1]{#2}}}

\def\round[#1]#2{\SI[round-mode=places,round-precision=#1,round-integer-to-decimal]{#2}{\noop}}

\def\rgetarr[#1]#2{\ifthenelse{\equal{\getarr[#1]{#2}}{ }}{}{\roundprefix[2]{\getarr[#1]{#2}}}}
\def\ngetarr[#1]#2{\ifthenelse{\equal{\getarr[#1]{#2}}{ }}{}{\SI[group-separator={\,}, group-minimum-digits=4]{\getarr[#1]{#2}}{\noop}}}

\def\calcpercprec#1#2#3{$\sim$\SI[round-mode=places,round-precision=#3]{\fpeval{((#1)*1.0)/((#2)*1.0)*100}}{\percent}}

\def\calcpercprec[#1]#2#3{$\sim$\SI[round-mode=places,round-precision=#1]{\fpeval{((#2)*1.0)/((#3)*1.0)*100}}{\percent}}
\def\calcpercprecnosim[#1]#2#3{\SI[round-mode=places,round-precision=#1]{\fpeval{((#2)*1.0)/((#3)*1.0)*100}}{\percent}}

\DeclareAcronym{SCHEDULING}{
  long        = CPU contention,
  short       = CPU contention,
  short-plural-form = CPU Contention,
  long-plural-form = CPU Contention,
  first-style  = long
}

\DeclareAcronym{BBR}{
  short        = BBR,
  long         = Bottleneck Bandwith and Round-trip Propagation Time,
  first-style  = long-short
}

\DeclareAcronym{BDP}{
  short        = BDP,
  long         = Bandwith Delay Product,
  first-style  = long-short
}

\DeclareAcronym{bw}{
  short        = BtlBw,
  long         = bottleneck bandwidth,
  first-style  = long-short
}

\DeclareAcronym{CCA}{
  short        = CCA,
  long         = congestion control algorithm,
  first-style  = long-short
}

\DeclareAcronym{cwnd}{
  short        = cwnd,
  long         = congestion window,
  first-style  = long-short
}

\DeclareAcronym{ECN}{
  short        = ECN,
  long         = Explicit Congestion Notification,
  first-style  = long-short
}

\DeclareAcronym{e2e}{
  short        = e2e,
  long         = end-to-end,
  first-style  = long-short
}

\DeclareAcronym{rtprop}{
  short        = RTprop,
  long         = round-trip propagation time,
  first-style  = long-short
}

\DeclareAcronym{RTT}{
  short        = RTT,
  long         = round-trip time
}

\DeclareAcronym{TCP}{
  short        = TCP,
  long         = Transmission Control Protocol
}

\DeclareAcronym{TSO}{
  short        = TSO,
  long         = TCP Segmentation Offload
}

\DeclareAcronym{VM}{
  short        = VM,
  long         = Virtual Machine
}

\begin{document}

\usetikzlibrary{positioning}
\usetikzlibrary{calc}

\title{2BRobust - Mitigating TCP BBR Performance Degradation in Virtual Machines under CPU Contention}

\author{Kathrin Elmenhorst}
\email{kelmenhorst@uos.de}
\orcid{0009-0000-0982-6456}
\affiliation{%
  \institution{Osnabrück University}
  \city{Osnabrück}
  \country{Germany}
}

\author{Nils Aschenbruck}
\email{aschenbruck@uos.de}
\orcid{0000-0002-5861-8896}
\affiliation{%
  \institution{Osnabrück University}
  \city{Osnabrück}
  \country{Germany}
}

\begin{abstract}
    Motivated by the recent introduction and large-scale deployment of BBR congestion control algorithms, multiple studies have investigated the performance and fairness implications of this shift from loss-based to delay-based congestion control.
    Given the potential Internet-wide adoption of BBR, we must also consider its \textit{robustness} in network and system scenarios. 
    One such scenario is Cloud-based Virtual Machine (VM) networking - highly relevant in today's CDN-centric Internet.
    Interestingly, previous work has shown significant performance problems of BBRv1 running in Xen VMs, with BBR performance dropping to almost zero when CPU credit is low.
    In this paper, we develop a framework for measuring TCP throughput under fully controlled CPU contention, which uses Linux deadline scheduling to emulate generalized \ac{SCHEDULING} conditions.
    Our measurements reveal that - in stark contrast to Cubic! - BBR throughput can break down during \ac{SCHEDULING} under any hypervisor and all tested BDP conditions.
    Characterizing this performance degradation on a fine-granular level, we show that CPU-limited BBR senders are capped at very low throughput levels below 10-20 Mbps.
    This finding implies that an Internet-wide shift from Cubic to BBR could harm the Internet's overall robustness, if not deployed with caution.
    To detect and overcome CPU-limited throughput, we propose a BBR patch which detects the problematic situation by monitoring inflight bytes and reacts by increasing the pacing rate to make better use of the available CPU time. We show that our BBR patch mitigates the throughput problem for the most critical cases.
\end{abstract}

% \ccsdesc[500]{Networks~Transport protocols}
% \ccsdesc[300]{Networks~Cloud computing}
% \ccsdesc[300]{Networks~Network performance evaluation}

% \keywords{BBR, TCP, Cloud, Virtual Machine}

\maketitle

\section{Introduction}

% prominence of BBR
TCP BBR \acp{CCA} have been around for almost a decade now, and have grown to largely impact the Internet's transport.
Their deployment has been accelerated due to the integration of BBRv1 into mainline Linux kernels, and the large traffic share of BBR's inventor Google who in 2023 reported that BBRv3 is used for all external google.com TCP traffic and tested for youtube~\cite{cardwell_bbr3_fixes2023}.
Looking over the TCP horizon, prominent QUIC implementations such as Google Chromium, Cloudflare quiche, and Facebook mvfst implement BBR versions on the application layer and thus further increase its presence on the public Internet.
In 2019, BBR was estimated to be used by 40\% of all downstram traffic~\cite{cc_census_mishra_2019}.
This shift from previous default TCP flavours (\eg Reno, Cubic) towards BBR represents a drastic change on the transport layer because BBR is based on an entirely different understanding of congestion: 
Instead of solely reacting to packet loss as a sign of congestion, BBR estimates the available bandwidth based on delivery rate and packet delay, and uses pacing by default.
Whilst BBR has shown to better use the available bandwidth in many scenarios~\cite{bbr_concept,cardwell_bbr3_fixes2023}, the paradigm shift requires a thorough evaluation.

% importance of testing, fairness (already lots of work), now: robustness, considering different use cases (VM)
In recent years, many studies have investigated the fairness of BBR, especially when coexisting with Reno or Cubic flows~\cite{cao2019,bbr_mobiles_2022,datta2023,gomez2024,zeynali2024,bbr_bless_2025}.
Besides fairness, another important aspect so far mainly overlooked is the \textit{robustness} of BBR in different usage scenarios, such as Cloud infrastructures.

\begin{figure}[t]
    \centering
    \includegraphics[width=\linewidth,trim={0 0.1cm 0 0},clip]{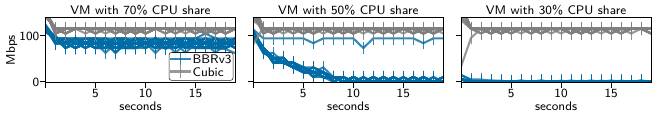}
    \vspace*{-0.5cm}
    \caption{Degrading BBRv3 throughput at decreasing VM CPU shares - Cubic robust. Experimental setup: \pref{sec:method}.\vspace*{-0.5cm} }
    \label{fig:motivating}
\end{figure}

% problematic finding: potentially not robust in VMs
In today's Internet, Cloud-based hosting is a specialized, yet prevalent usage scenario for BBR:
CPU and network resources can be limited during peak usage hours, and when CPU contention occurs, networking in virtualized environments is typically more affected than on CPU-constrained bare-metal systems, because every part of packet transmission, including tasks offloaded to the virtualized network card, depend on the host's CPU.
Thus, virtualization could have unexpected effects on BBR's bandwidth estimation algorithm which is certainly more complex (and more fragile?) than traditional loss-based algorithms.
Interestingly, previous work~\cite{ha2021, elmenhorst2025} has measured serious throughput problems when running BBR in \acp{VM} under CPU contention. 
\pref{fig:motivating} shows an example for degrading BBR throughput in \acp{VM} with 30, 50, and 70\% CPU shares.

Based on these initial findings, in this paper we set out to better measure, understand, and overcome the problematic throughput degradation, independently of specific \ac{VM} hypervisors, and under varying sender conditions.
Our contributions, as presented in \pref{sec:method}-\ref{sec:solution}, are as follows:
\begin{enumerate}[label=\circlelabel{\Roman{enumi}}{L\arabic{enumi}}]
    \item \textbf{Controlled \ac{SCHEDULING} testbed}: We develop a deadline scheduling-based method to measure BBR throughput in \acp{VM} during emulated \ac{SCHEDULING} on real systems.
    Our method can not only reproduce the performance degradation seen in prior work, but allows for precise control over the \ac{SCHEDULING} and is independent of any specific hypervisor. For validation, the emulation is complemented by a real hypervisor scheduling setup.
    \item \textbf{Profile of CPU-limited sockets}: We evaluate BBR state and throughput during multiple \ac{SCHEDULING} and network scenarios, and so identify the (not so) specific conditions triggering the problematic degradation. We use Cubic as a loss-based reference, and confirm that the BBR degradation occurs in all tested kernels, \acp{VM}, and across BBR versions.
    \item \textbf{Solution approach}: For mitigation, we propose a BBR patch which uses a BDP-relative bytes-inflight metric to detect CPU-limited sockets and reacts by conservatively increasing the pacing. We show that patched BBR can successfully shift the problem boundary.
\end{enumerate}
% --- overlay arrows between the circles ---
\begin{tikzpicture}[remember picture,overlay,>=stealth]
  \draw[->] (sL1) -- (L2);
  \draw[->] (sL2) -- (L3);
\end{tikzpicture}

\section{BBR Congestion Control Algorithms}
\label{sec:background}
In this section, we will elaborate on the inner workings of BBR by (1) explaining how the \ac{CCA} models the link and regulates the sending rate, (2) describing the different algorithm phases, and finally (3) outlining the main differences between the existing BBR versions.

\subsection{BBR Model}
\label{sub:bbr_model}
\ac{BBR} is a \ac{CCA} first introduced by Google in 2016~\cite{bbr_concept}. 
Most prior \acp{CCA}, including the major TCP versions based on Reno and Cubic, react to loss as a sign of congestion, reasoning that congested links drop packets when buffers overflow.
In contrast, \textit{latency-based} \ac{BBR} tries to send at a rate where the bottleneck bandwidth is fully utilized, but buffers are still relatively empty.
This approach is meant to proactively avoid congestion and keep latencies low.

To this end, BBR monitors \acp{RTT} and delivery rates (acked bytes / time) to estimate the \ac{bw} and \ac{rtprop} of the connection.
Scaled by phase-dependent gain factors, \ac{bw} and \ac{rtprop} directly regulate the three driving forces for the sending rate, namely the congestion window, the pacing rate, and the TSO burst size~\cite{cardwell-iccrg-bbr-congestion-control-02}.
As elaborated in the following, the pacing rate determines the sending rate (speed), while the congestion window is the absolute limit for inflight bytes per RTT.

\pgfdeclarelayer{background}
\pgfsetlayers{background,main}
\begin{figure}[t]
  \captionsetup[subfigure]{labelformat=simple,labelsep=space}
\renewcommand\thesubfigure{Fig. \thefigure\alph{subfigure}}
\begin{subfigure}{0.54\linewidth}
    \begin{tikzpicture}
    % sender
    \node (box1) [draw, rectangle, minimum width=0.25cm, minimum height=1cm, fill=gray] {};
    \node (box1_text) [left=0cm of box1] {Send};
    % receiver
    \node (box2) [draw, rectangle,  minimum width=0.25cm, minimum height=1cm, right=5cm of box1, fill=gray] {};
    \node (box2_text) [right=0cm of box2] {Recv};
    \node (box5) [fill, rectangle, minimum width=4cm, minimum height=0.5cm, align=center, fill=lightblue, opacity=0.5, above=-0.5cm of $(box1.north)!0.5!(box2.north)$] {};
    \node (box6) [fill, rectangle, minimum width=4cm, minimum height=0.5cm, fill=lightblue, opacity=0.5, below=-0.02cm of box5] {};
    \begin{pgfonlayer}{background}
    \node (ellipse1) [draw, ellipse, fill=red, minimum height=0.5cm, minimum width=0.25cm, left=-0.2cm of box5, label=above:{\footnotesize \textcolor{red}{BtlnBW}}] {};
    \node (ellipse2) [draw, ellipse, fill=white, minimum height=0.5cm, minimum width=0.25cm, right=-0.2cm of box5] {};
    \node (ellipse3) [draw, ellipse, fill=white, minimum height=0.5cm, minimum width=0.25cm, left=-0.2cm of box6] {};
    \node (ellipse4) [draw, ellipse, fill=white, minimum height=0.5cm, minimum width=0.25cm, right=-0.2cm of box6] {};
    \end{pgfonlayer}
    \draw[-] (ellipse1.north) -- (ellipse2.north);
    \draw[-, thick, red] (ellipse3.south) -- node[below] {\footnotesize RTT/2} (ellipse4.south);
    \draw[<-, thick, darkerblue] (ellipse1.west) -- (ellipse1-|box1.east);
    \draw[->, thick, darkerblue] (ellipse3.west) -- (ellipse3-|box1.east);
    \draw[->, thick, darkerblue]
      (ellipse2.center) 
        .. controls +(0.9,0) and +(0.9,0) ..
      (ellipse4.center);
    \node at (2.75cm, 0.05cm) {\footnotesize {\color{darkerblue} \textbf{BtlnBW x RTT}}};
    \end{tikzpicture}
    \caption{Pipe-model of the Bandwidth Delay Product (BDP).}
    \label{fig:bdp_pipe}
\end{subfigure}
\begin{subfigure}{0.45\linewidth}
    \centering
    \begin{tikzpicture}
    \node[draw=none,fill=none] at (1.9,1.3){\includegraphics[width=0.8\linewidth]{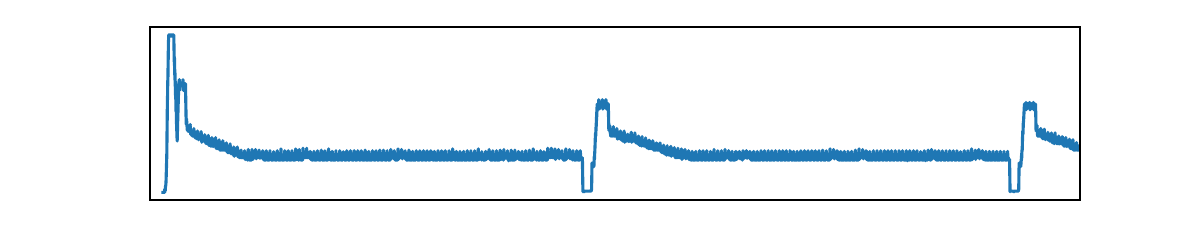}};
    \node[draw, circle,minimum width=0.5cm, label=above:{\vphantom{Drain}\footnotesize Startup}] (A) at (0,0)  {};
    \node[draw, circle,minimum width=0.5cm, label=above:{\vphantom{Startup}\footnotesize Drain}] (B) at (1.25,0)  {};
    \node[draw, circle, minimum width=0.5cm, label=above:{\vphantom{Startup}\footnotesize ProbeBW}] (C) at (2.5,0)  {};
    \node[draw, circle, minimum width=0.5cm, label=above:{\vphantom{Startup}\footnotesize ProbeRTT}] (D) at (3.75,0)  {};
    \path (A) edge [->, thick, darkerblue] (B);
    \path (B) edge [->, thick, darkerblue] (C);
    \path (C) edge [thick, darkerblue, loop below] (C);
    % \path (C) edge [->, thick, darkerblue] (A);
    \path (C) edge [->, thick, darkerblue, bend left] (D);
    \path (D) edge [->, thick, darkerblue, bend left] (C);
    \draw[-, red] (0,0.65) -- (0.1,1);
    \draw[-, red] (1,0.7) -- (0.2,1);
    \draw[-, red] (2.3,0.7) -- (2.45,1);
    % \draw[->, red] (3,0.7) -- (3.8,1);
    \draw[-, red] (3.5,0.7) -- (3.65,1);
    \end{tikzpicture}
    \caption{Simplified BBR State Model~\cite{bbr3}.}
    \label{fig:bbr_state_model}
\end{subfigure}
\end{figure}

\subsubsection*{\Ac{cwnd}.}
As for any TCP connection, the \ac{cwnd} limits the amount of data inflight, \ie sent but not yet acknowledged.
BBR sizes the \ac{cwnd} to the estimated \ac{BDP} multiplied by a gain factor, \ie \ac{bw} $\cdot$ RTprop $\cdot$ cwnd\_gain.

In general, a TCP sender can only fully utilize the bandwidth if there is at least 1 BDP of data inflight~\cite{understanding_bbr_2018}.
The 1 \ac{BDP} threshold known as Kleinrock's optimal point of operation can be illustrated by thinking about the TCP connection as a pipe where the cross-sectional area is the bottleneck bandwidth and the length is the RTT (\subref{fig:bdp_pipe}).
BBR mostly uses cwnd\_gain $= 2$, \ie \ac{cwnd} = 2 BDP, accounting for potential retransmissions which delay the acknowledgement by an \ac{RTT}.

\subsubsection*{Pacing rate.}
Instead of sending as fast as possible until the \ac{cwnd} limit is reached, BBR applies a pacing rate to limit burtiness and thus avoid (temporary) congestion at bottleneck buffers.
Choosing a pacing rate close to the estimated bandwidth ensures that the \ac{cwnd} is evenly spread out over the \ac{RTT}, \ie there is a small, constant timing offset between subsequent packets~\cite{aggarwal_pacing2000}.
BBR calculates this inter-packet offset as \textit{packet size} / pacing rate.
Since Linux TCP commonly uses TCP Segmentation Offloading (TSO), \textit{packet size} is equal to the TSO burst size which, as described next, is set so that the effective send interval is 1 millisecond at most data rates.

\subsubsection*{TSO burst size.}
In the Linux kernel, pacing is closely related to TSO which is a mechanism to reduce per-packet CPU processing overhead: 
the kernel passes oversized TCP segments, or "TSO bursts", to the NIC which takes on the task of splitting the large segment into MTU-complying TCP segments~\cite{pacing_2025}.
BBR explicitly sets the TSO burst size~\cite{cardwell-iccrg-bbr-congestion-control-02} as 1 millisecond of data at the current pacing rate ($BBR.pacing\_rate \cdot 1ms$), but at least 2 packets, and at most 64 KB. For data rates under 1.2 Mbps, BBR completely disables TSO, \ie sets burst site to 1.
With the pacing interval definition above, TCP BBR sends a TSO burst \textit{every 1 ms} for pacing rates under 512 Mbps, and more often for pacing rates over this value.

\subsection{BBR Phases}
Throughout a connection's lifetime, BBR goes through different phases, as simplified in \subref{fig:bbr_state_model}.

\paragraph{Startup and Drain}
Similar to "slow" start in TCP Reno and Cubic, a new BBR connection doubles the \ac{cwnd} and pacing rate each \ac{RTT}~\cite{bbr_concept} to quickly approach the bandwidth limit.
When the delivery rate stalls, BBR exits Startup and momentarily decreases the sending speed, \ie pacing rate, to drain built-up queues.
During this phase, cwnd\_gain and pacing\_gain are at least 2, and BBR collects delivery rate samples to estimate the current \ac{bw}.
For the remainder of the connection, BBR alternates between ProbeBW and ProbeRTT.

\paragraph{ProbeBW}
Most of a BBR connection's time is spent in ProbeBW.
This phase is characterized by a close approximation of the available bandwidth, where BBR is always trying to increase throughput in case more bandwidth becomes available.
During this phase, cwnd\_gain is 2, and pacing\_gain between 0.75 and 1.25, and BBR collects delivery rate samples to estimate the current \ac{bw}.

\paragraph{ProbeRTT}
Since some (non-excessive) queue built-up is to be expected when sending at the maximum available bandwidth and \ac{cwnd} of 2 \ac{BDP}, the measured \ac{RTT} will include queuing delay during ProbeBW. 
To measure the propagation delay only, BBR tries to drain queues by shortly reducing \ac{cwnd} and pacing rate.
The interval between subsequent ProbeRTT phases varies between BBR versions, as described below, but is generally in the range of a few seconds.
During this phase, cwnd\_gain and pacing\_gain are $\leq$1, and BBR collects \ac{RTT} samples to estimate the current \ac{rtprop}.

\subsection{BBR Versions}
BBRv1 was first published in 2016, and has been integrated into mainline Linux from kernel 4.8.
Since then, two more versions have been developed to overcome issues identified in the first version.

Importantly, while BBRv1 is indifferent to loss signals, BBRv2 and BBRv3 respond to loss and \ac{ECN} signaling, so that retransmissions can be reduced~\cite{datta2023}.
Another change after version 1 added recently measured inflight samples as a model parameter to control the cwnd.
In BBRv1, bandwidth probing (ProbeBW) oscillates around the current pacing rate via a fixed list of pacing\_gain values, and is interrupted by ProbeRTT every 10 seconds.
In contrast, BBRv2 and BBRv3 are more adaptive to competing flows in ProbeBW due to less fixed pacing\_gain cycles. 
The newer versions enter ProbeRTT more frequently, \ie every 5 seconds, but using a less drastic cwnd reduction than BBRv1.

The changes from BBRv2 to BBRv3 are reportedly~\cite{cardwell_bbr3_fixes2023} rather incremental, even though fairness measurements saw similar aggressiveness between BBRv1 and BBRv3~\cite{zeynali2024}.
The fixes mainly revolve around re-parametrizing ProbeBW, as detailed in related work~\cite{zeynali2024}.

\section{Related Work}
\label{sec:related}

BBR initially sparked the interest of the academic research community after BBRv1 was first introduced in 2016, and has continuously motivated new studies, especially when new versions of the algorithm were released. 
This section gives an overview over existing papers on BBRv1-v3 performance evaluations and, secondly, specifically on BBR in \acp{VM}.

There is a large amount of previous work studying BBR's performance and fairness implications under different scenarios.
Methodically, existing works can be classified into measurement studies~\cite{cao2019,bbr_mobiles_2022,datta2023,gomez2024,zeynali2024,bbr_bless_2025} evaluating BBR performance under a variety of different network and traffic pattern conditions on the one hand, and on the other hand model-driven research~\cite{bbr_ware_2019,bbr_mishra_2022,bbr_scherrer_2022} where BBR properties like throughput and fairness are analytically modelled and validated via experiments.
Important findings include that BBR lacks fairness against loss-based flows, primarily in versions 1 and 3~\cite{cao2019,zeynali2024}, and that BBR works best under shallow bottleneck buffer conditions, while loss-based \acp{CCA} like Cubic are advantageous in networks with deep buffers~\cite{datta2023}.

These foundational studies on general BBR performance give way to more detailed evaluations for specific use cases, such as Cloud networking and \ac{VM}.
Recently, Phanekham et al.~\cite{cloud_phanekham_2025} evaluated throughput in Cloud networking scenarios, that is, they performed throughput measurements between Google Cloud \acp{VM} located at different sites.
\textit{Using dedicated processors for each VM}, they find that BBRv1 and v2 outperform loss-based \acp{CCA} especially on lossy high-latency paths.

Regarding the more narrow scope of this paper, there are two previous publications pointing out the degradation of BBR in \acp{VM} under \ac{SCHEDULING} conditions:
Ha et al.~\cite{ha2021} first theoretically analyzed the effect of VM \ac{SCHEDULING} on a simplified version of the BBR ProbeBW algorithm. 
Practically, the authors measure performance degradation of BBRv1 and v2 in \acp{VM} on custom and AWS Xen-based hypervisors.
They model \ac{SCHEDULING} severity via Xen weights and AWS credit balances as proxy metrics, arguing that Xen hypervisors allocate higher CPU shares for \acp{VM} with higher weight/credit.
Their measurements on an AWS EC2 \ac{VM} show that a credit balance of 0.15-0.2 leads to throughput close to zero.
In a testbed with a custom-configured Xen hypervisor, BBR throughput drops to around 5 Mbps when the \ac{VM} has a credit of 64 - which they define as "heavy scheduling" conditions.
Complementary Cubic measurements are not provided for reference.

To combat the BBR problem, Ha et al. propose a patch which uses individual packet timestamps to detect off-CPU times and increases BBR's bandwidth estimate, pacing rate, and congestion window according to the \ac{VM}'s current estimated CPU share. 
While this approach shows improved throughput in most Xen scheduling scenarios, it stretches protocol responsibilities by requiring BBR to compute the current CPU share of the \ac{VM} guest.
Furthermore, the fairness implications of over-estimating the available bandwidth and increasing the congestion window gain might be problematic, especially when considering the case where \ac{SCHEDULING} conditions end abruptly.
Since the patch was not released publicly and we were not able to reproduce their measurement framework, we cannot provide a sound comparison between their patch and our solution (\pref{sec:solution}).

More recently, Elmenhorst and Aschenbruck~\cite{elmenhorst2025} measured significant BBR throughput degradation in VMs under suboptimal \ac{VM} scheduling, underlining robustness disadvantages in BBR compared to Cubic.\newline

Realizing both the high impact, but also the limiting methodology of previous findings, we aim to isolate the impact of \ac{SCHEDULING} on BBR state and throughput - independent of the VM platform - and compare it to Cubic under varying path-\acp{BDP}. 
To this end, our first contribution (I) is to emulate \textit{controlled} \ac{SCHEDULING} for a TCP BBR testbed, as described in the following.

\section[Controlled \acp{SCHEDULING} Testbed]{Controlled \acp{SCHEDULING} Testbed \hspace{0.25em} \texorpdfstring{\ISymbol}{}}
\label{sec:method}

Our measurement and evaluation setup\footnote{Scripts available under: \url{https://github.com/sys-uos/bbr-cpu-contention}} relies on two basic building blocks: 1) A TCP performance test setup, and 2) the modeling and emulation of variable \ac{SCHEDULING} conditions.

\subsection{TCP Measurement Setup}
\label{sub:tcp_setup}

\begin{figure}[t]
\centering
\includegraphics[width=0.89\linewidth]{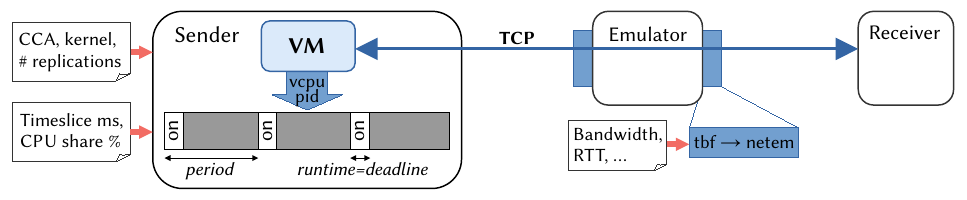}
\caption{Evaluation framework using a simple TCP Setup with link emulation, and CPU contention modeled as deadline scheduling of the sender VM's vCPU.}
\label{fig:method}
\end{figure}

Our TCP measurement setup, as shown in \pref{fig:method}, consists of two physical Dell PowerEdge R620 servers A and B, and an intermediate network emulator, connected via 25 Gbps Ethernet (Broadcom BCM57414 NetXtreme-E 25Gb).
The servers run Ubuntu 24.04 with Linux 6.8, and are equipped with 200 GB RAM and 24 physical CPU cores.
The network emulator bridges (layer 2) packets between A and B. 
Using the bottleneck link model which argues that, from the perspective of TCP, a multi-link network connection behaves like a single link with the accumulated delay and bottleneck bandwidth~\cite{bbr_concept}, we emulate different link conditions by configuring Linux tc netem and token buckets at the bridge interfaces, as detailed below.

We deploy QEMU \acp{VM} on the sender host using a KVM hypervisor, and configure them with 16 GB virtual RAM and bridged \textit{virtio} network adapters to connect to the host network.
The guests run Debian 12 with kernel v6.1 when running Cubic and BBRv1.
As BBRv2 and BBRv3 have not yet been upstreamed to mainline, we use dedicated Google custom kernels based on Linux v5.13~\cite{bbr2} and v6.13~\cite{bbr3} - the only official sources of TCP BBRv2 and BBRv3.

\paragraph{Network parametrization and measurement}
To parametrize our network emulation and host sockets, we start off with a set of fixed parameters which are kept constant throughout the main experiments:
For the main evaluation, we set the network bottleneck buffer size to 1 \ac{BDP} and the packet loss rate to 0\%.
% netem seed parameter for reproducibility
We enable \ac{TSO} on both the sender and the emulator node, and set the sender's maximum socket buffer size to the highest allowed value (2147483647) so that it does not become the connection bottleneck.
Based on this fixed parametrization, we vary the \ac{RTT} between 10ms and 40ms, and the maximum bandwidth between 10 and 1000 Mbps.
These levels are meant to resemble "realistic" \ac{BDP} conditions in wired Internet settings.

For each individual TCP BBR measurement, we parametrize the network emulation and start a 20-second iperf3 bulk transfer with specified sender \ac{CCA}.
We repeat each run at least 30 times.

\paragraph{Setup validation}
For any tested bandwidth, we successfully validate that we can saturate the link between the two endpoint VMs using UDP, confirming that our setup itself does not contain the bottleneck of the connection.
We further confirm that TCP performance is not inhibited when the VM has exclusive usage of the host resources, \ie without \ac{SCHEDULING}.
Throughout the evaluation, we cross-check whether \ac{CCA}-correlated effects are actually kernel-specific - since BBRv2 and v3 are not upstreamed to mainline Linux - but do not find such a case.

\subsection{Emulating \acp{SCHEDULING} via Deadline Scheduling}
\label{sub:scheduling_model}
As elaborated in the previous section, one challenge of previous work is the lack of a \textit{controllable}, generalizable evaluation architecture for conducting throughput measurements under repeatable \ac{SCHEDULING} and network conditions.
This poses a challenge both for Cloud-based measurements in the wild, and when using self-hosted instances of common hypervisors:
In the wild, \eg when measuring in \acp{VM} hosted by Cloud providers, researchers usually cannot control host-side scheduling and CPU resource utilization, and even lack data access to correlate CPU events with the throughput achieved in the VM guest.
The measured results are specific to the measurement time, as well as the \ac{VM} and hypervisor configurations.
When running a dedicated instance of a common hypervisor, \eg Xen, different patterns of \ac{VM} CPU time sharing can only be controlled via abstract proxy parameters, \ie "weight"~\cite{ha2021}, and the hypervisor's custom scheduling algorithm adds complexity and hinders generalizability.

To overcome these issues, we develop a simple, yet effective method to emulate variable, repeatable \ac{SCHEDULING} conditions in high resolution.
Due to the generality of the approach, our \textit{absolute} results are not equivalent to any specific hypervisor model, but the effects generated in the emulation are applicable to every \ac{SCHEDULING} setting.

The key mechanism is that \textbf{we force precise timeslices on the sender \ac{VM} by means of Linux \textbf{deadline scheduling}}~\cite{kernel_deadline}.
Deadline scheduling is a Linux scheduling algorithm for real-time tasks based on the Earliest Deadline First (EDF) algorithm.
As shown in \pref{fig:method}, it enforces periodic on- and off-CPU times controlled by three parameters, namely the process' \textit{runtime}, the length of one full \textit{period}, and the time at which the process must have finished its work within each cycle (\textit{deadline}, which we set equal to \textit{runtime} for full control).
In the following, we refer to the runtime as the "timeslice length", and use the term "CPU share" for the \%-ratio between runtime and period.

To implement this model on the sender host, we change the scheduling policy of the processes running the \ac{VM}'s virtual cores to SCHED\_DEADLINE, varying the timeslice length between 1ms and 20ms, and choosing period lengths accordingly so that the CPU shares are between 10\% and 70\%.
Since the virtual core processes are the only real-time (RT) tasks on the host system, the deadline scheduling is enforced before any other system task can be scheduled.

Whilst this deadline scheduling approach cannot mirror the complexity of real-world hypervisors, it allows for fine-grained control over the emulated \ac{SCHEDULING} parameters, in particular the \ac{VM}'s timeslice length and CPU share, and provides generalizability precisely because it reduces scheduling to its most basic parameters.
In \pref{sec:real_kvm} the measurements from our CPU emulation will be validated using real KVM hypervisor behavior.

\section[Profile of CPU-Limited Sockets]{Profile of CPU-Limited Sockets \hspace{0.25em} \texorpdfstring{\IISymbol}{}}
\label{sec:degradation}
In this section, we will use the deadline scheduling-based measurement framework described above to run and evaluate BBR throughput measurements under variable \ac{SCHEDULING} and network conditions.
Starting off with an overview and root cause analysis of the general problem in \pref{sub:eval_overview}, we then evaluate the parameter influence of timeslice lengths, bandwidth, and RTT in more detail.

\subsection{Problem Overview and Underlying BBR State}
\label{sub:eval_overview}

\begin{figure}[t]
    \centering
    \includegraphics[width=\linewidth]{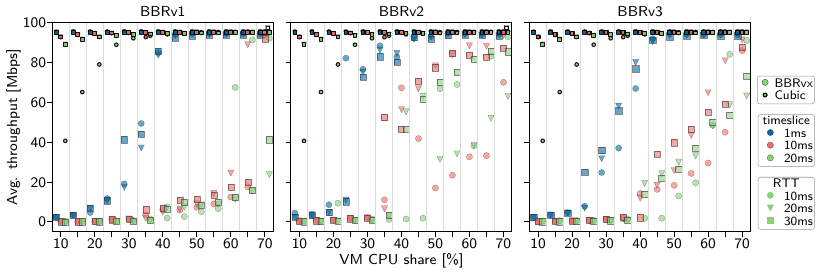}
    \caption{BBR vs. Cubic throughput (medians) for varying \ac{SCHEDULING} conditions and \acp{RTT}. Bottleneck bandwidth is fixed to 100 Mbps.}
    \label{fig:first_eval}
\end{figure}

Even at first glance, \pref{fig:first_eval} shows that BBR throughput in \acp{VM} is heavily inhibited by \ac{SCHEDULING}, whilst Cubic stays robust.
The plot shows data rates for a 100 Mbps link with varying \acp{RTT}, timeslice lengths, and CPU shares.
In this aggregated view, each data point corresponds to the median of 30 samples, where each sample is the average throughput of a 20-sec. TCP connection.

Considering Cubic as a reference, the results are quite dramatic: 
While \textbf{Cubic reaches the maximum bandwidth of 100 Mbps in nearly all tested scenarios, BBR throughput is heavily inhibited by \ac{SCHEDULING}} across all tested algorithm versions and configured delays.
Though the impact is notably smaller in BBRv2, no BBR version exceeds 11 Mbps in the median when the \ac{VM} receives 10-25\% CPU shares, regardless of the timeslice lengths.
In case of the shortest timeslice (1 ms), BBRv1 and v3 only start approaching the full bandwidth at a 45\% CPU share, for longer timeslices they need at least a 65\% CPU share to exceed 80\% of the bandwidth in the median.

This overview raises multiple questions: 
Why is the BBR algorithm so much more affected by \ac{SCHEDULING}?
Why does BBRv2 show a slightly different trend than versions 1 and 3?
And what is the (interdependent) impact of varying timeslice lengths and network parameters on the performance?
While the latter question will be addressed in \pref{sub:eval_network}, 
we first aim to better understand the general performance-limiting problem (und thus, questions 1-2) by investigating BBR throughput on two more levels of detail, zooming in to better grasp the underlying dynamics.
\newline

\begin{figure}
    \centering
    \includegraphics[width=\linewidth]{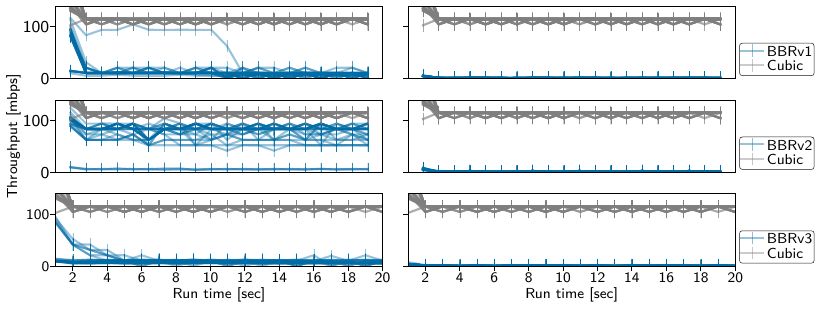}
    \caption{Throughput over time for BBRv1-3 compared to Cubic, with 25\% CPU shares, at 100 Mbps, 10 ms RTT. Left: \textbf{1}ms timeslice, right: \textbf{10}ms timeslice.}
    \label{fig:throughput_timeseries}
\end{figure}

\paragraph{Zoom-Level 1: Throughput dynamics over connection time}
To see time-dependent features hidden in the aggregated view, we now zoom-in and analyze the throughput \textit{over time}.
\pref{fig:throughput_timeseries} shows time series results for different \acp{CCA}, and timeslice lengths.
The shown comparison of the two configured timeslice lengths (1, 10ms) implies that longer timeslices (right) have a stronger impact on BBR, even when the relative CPU share is the same. 
In this case, no BBR version ever exceeds 5 Mbps, while Cubic still consistently achieves the 100 Mbps bandwidth.

With shorter, \ie 1 ms, timeslices (\pref{fig:throughput_timeseries}, left), many BBRv1 runs achieve rates close to the available bandwidth in Startup, but then quickly drop below 11 Mbps.
Another fraction of the BBRv1 runs, as well as most BBRv3 runs, never go up to the available bandwidth at all, similar to the 10ms timeslice case.
For BBRv2, the picture looks rather different:
After a short initial throughput drop, most test runs exhibit rates from 50 Mbps to 100 Mbps, \ie the configured bandwidth limit.
Still, the variance is much higher than for Cubic, and 10\% of the test runs fail to reach the bandwidth in Startup completely.
This seems to suggest that \textit{if} BBRv2 manages to reach the bandwidth limit during Startup despite the \ac{SCHEDULING}, it can mostly maintain this rate - in contrast to BBRv1 and v3 which drop to low rates even if Startup was successful.

\paragraph{Zoom-Level 2: BBR State Analysis}
For an even more detailed analysis and understanding of the underlying algorithm behavior, we now analyze BBR state variables during the connection.
This information can be extracted from the TCP socket manually on a custom server (\texttt{TCP\_INFO}, \texttt{TCP\_CC\_INFO}), or by means of standard Linux utilities like the \texttt{ss}-tool.
Note that both methods add processing load on the VM and could thus slightly skew the throughput results, so we run the tests for this section separately from the other measurements.
For better replicability and integration with our iperf3-based measurements, we use the \texttt{ss}-tool.

\begin{figure}[t]
    \centering
    \includegraphics[width=\linewidth]{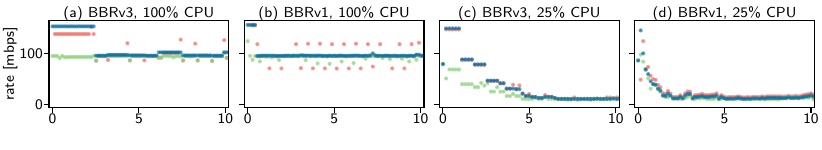}

    \vspace*{-0.2cm}
    
    \includegraphics[width=\linewidth]{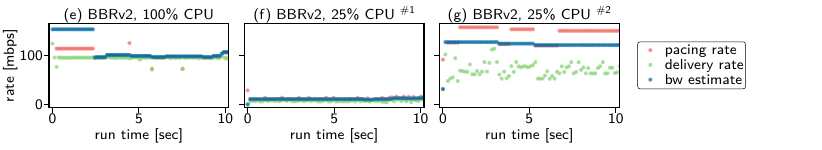}
    \caption{Pacing rate, delivery rate, and BBR bandwidth estimate of individual flow samples for different BBR versions and CPU shares.}
    \label{fig:zoom3}
\end{figure}

For individual representative samples, \pref{fig:zoom3} shows BBR bandwidth estimates, as well as TCP delivery and pacing rates when using different \acp{CCA} and \ac{SCHEDULING} conditions.
Since the previous zoom levels have shown a notable difference between BBRv1 and v3 on the one hand, and BBRv2 on the other hand, we will analyze them separately.

For BBRv1 and v3, that is \pref{fig:zoom3} a) - d), we observe the following: 
Initially, BBR's high pacing gain during the Startup phase leads to a high pacing rate overshooting the bandwidth limit.
With this pacing rate, the TCP connection can deliver bytes at a rate close to the limit and BBR's bandwidth estimate is adapted to this delivery rate. After Startup, \ie when BBR assumes that the bandwidth limit has been reached, the pacing gain is reduced to around 1, periodically probing for more bandwidth in ProbeBW.
If the \ac{VM} has unrestricted CPU access (100\%), the delivery rate (and thus bandwidth estimate) stays close to the limit after Startup. 
With \ac{SCHEDULING}, the delivery rate first goes up close to the bandwidth limit as well, but when the bandwidth estimate (and thus pacing rate) is adjusted after the initial overshoot during Startup, the delivery rate also decreases, now sending below the link limit.
As theoretically analyzed in prior work~\cite{ha2021} and practically shown here, this is because \textbf{VM-scheduled BBR needs a higher pacing rate to put enough data inflight while the VM can actually use the CPU}. In particular, TCP pacing uses fixed pacing intervals, \ie inter-packet timing offsets, even though the sparse CPU time situation would require to send packets faster and so better utilize the available timeslices.
While this should apply to all paced traffic, we see that paced Cubic, \ie Cubic with \texttt{fq qdisc}, is strikingly less affected by the problem (\cf \pref{fig:eval_cubic_fq}).
This is because BBR reacts counterproductively: Due to the variables' dependencies in BBR, the reduction in delivery rate causes a reduction of the bandwidth estimate, which again causes a reduction of the pacing rate which further reduces the delivery rate\footnote{The cwnd is also reduced during this cycle, as it also depends on the bandwidth estimate. However, since the cwnd is sized at 2 \ac{BDP} whilst the pacing rate aims at 1 BDP inflight, the latter should be the limiting factor to the throughput here.}.
This "downward spiral", which can be clearly seen in the plots, seems to continue until a lower throughput bound is reached.
In other instances (not shown), the delivery rate cannot keep up with the pacing rate already during Startup, so BBR immediately exits Startup, never even approaching the available bandwidth.

For BBRv2, the socket statistics uncover an unexpected behavior which explain the throughput differences during \ac{SCHEDULING}.
Note that due to the high variance in BBRv2 throughput samples during \ac{SCHEDULING}, we show three representative samples: one without \ac{SCHEDULING} (e), one where BBRv2 throughput is clearly inhibited by \ac{SCHEDULING} (f), and one where BBRv2 can send close to the available bandwidth regardless of \ac{SCHEDULING} (g).
While the case without \ac{SCHEDULING} seems to behave similarly to the other BBR versions, \pref{fig:zoom3} g) reveals a truly unexpected behavior: in contrast to our understanding of the BBR algorithm, the delivery rate - which, just as for BBRv1 and v3, fails to fully reach the bandwidth limit during \ac{SCHEDULING} - does not seem to really affect the bandwidth estimate.
In fact, the so-behaving BBRv2 runs have in common that after a certain point in time, the bandwidth estimate is \textit{never updated again}.
Since the fixing of a deprecated BBR version is out of scope for this paper, we do not further inquire exactly when and why this happens, however, Google's BBRv3 bug fix report mentions "circular dependence" between the bandwidth estimate and max in-flight data in v2~\cite{cardwell_bbr3_fixes2023} which could potentially contribute to the behavior seen in this scenario.
Ironically, this behavior makes BBRv2 less sensitive to \ac{SCHEDULING} in some cases, \ie when the bandwidth estimate "freezes" at a lucky high point and thus stops the downward spirale described above.
Due to this uncharacteristic behavior and the fact that both BBRv1 and v2 are deprecated, we will focus on the current BBRv3 in the remainder of this paper (additional plots for BBRv1 can be found in the Appendix \pref{fig:rate_diff_bbrv1}, \ref{fig:delay_diff_bbrv1} and \ref{fig:eval_solution_agg_bbrv1}).

\subsection{The (Interdependent) Impact of Network and \acp{SCHEDULING} Parameters}
\label{sub:eval_network}
We will now investigate how variable network parameters can interact with \ac{SCHEDULING} conditions, and thus influence the resulting throughput in a more-dimensional way.

\begin{figure}[t]
    \centering
    \includegraphics[width=\linewidth]{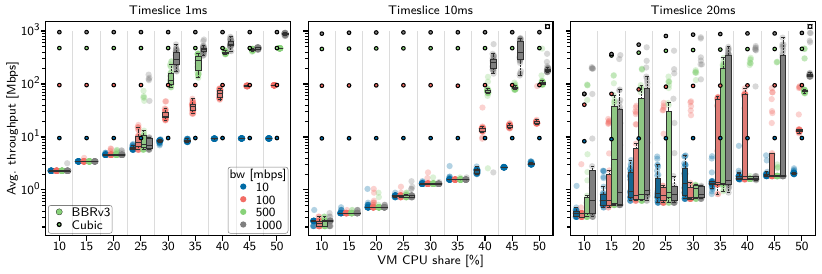}
    \caption{BBRv3 throughput at 10 ms RTT for varying bottleneck bandwidths. Note the logarithmic y-scaling.}
    \label{fig:rate_diff}
\end{figure}

\paragraph{Bottleneck bandwidth}
As shown in \pref{fig:rate_diff} for BBRv3 at 10 ms RTT, comparing 1-20 ms timeslices, all tested bandwidth limits lead to the same throughput if the CPU share is small.
In the 10 ms timeslice case, this threshold is at 35\% CPU share, where throughput is roughly capped at 20 Mbps.
After the threshold, the increment at each CPU-share step grows relatively to the bottleneck bandwidth.
When increasing the timeslice length to 20 ms, the variance is much higher and no strict upper bound exists.
However, the medians of the different bottleneck bandwidths are still approx. the same for CPU shares up to 45\%.

The minor impact of the bandwidth at low CPU shares can be explained based on Linux' pacing algorithm, as outlined in \pref{sec:background}: 
for the most part, faster pacing rates lead to larger TSO burst sizes, not smaller pacing intervals.
In particular, the offset between subsequent packets is constant at 1 ms, which means that the send \textit{timing} - which is affected by timeslicing - does not change for higher bandwidths.
As elaborated in \pref{sec:background}, this behavior only changes at very low bandwidths (below 1.2 Mbps), where TSO is disabled, and high bandwidths (over 512 Mbps) where the pacing interval becomes smaller than 1 ms~\cite{pacing_2025}, however, our results show that even at 1000 Mbps, BBR throughput is limited to the same absolute bounds.

Based on these insights, we classify BBR throughput during CPU contention as \textit{strictly CPU-limited} when the throughput is limited independently of bandwidth, recovering, or non-limited.

\begin{figure}[t]
    \centering
    \includegraphics[width=\linewidth]{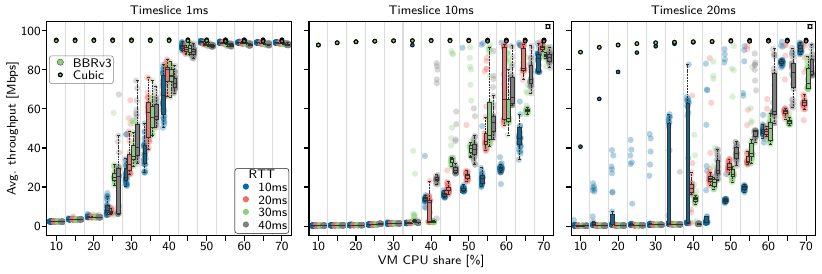}
    \caption{BBRv3 throughput at 100 Mbps bottleneck bandwidth for varying \acp{RTT}.}
    \label{fig:delay_diff}
\end{figure}

\paragraph{Timeslice length and RTT}
As shown in \pref{fig:first_eval}, longer timeslices inhibit BBR's sending capabilities more than shorter timeslices, when compared at the same relative CPU share. 
For example, BBR throughput recovers at 40-45\% CPU share with a 1 ms timeslice, but only around 65-70\% with 10 and 20 ms timeslices.
Interestingly, the 20 ms timeslice case also leads to strongly increased variance compared to the other timeslice lengths (\cf \pref{fig:overview_stddev}).

Given that this effect occurs especially for smaller \acp{RTT}, the increased variance is likely due to our choice of relatively short \acp{RTT}.
It is intuitive to assume an interdependency between the impact of the timeslice length and the RTT.
Consider for example that in the 1 ms timeslice case, the \ac{VM} can use the CPU at least once per \ac{RTT} for all configured \acp{RTT} and CPU shares, because even at 10\% CPU share the off-CPU time (9 ms) is not long enough to "swallow" an entire RTT.
Conversely, with 10-20 ms timeslices, the span of an entire \ac{RTT} can fall into an off-CPU interval which decreases the sender's capabilities to fill the pipe - in the extreme case, \ie 20 ms timeslice and 10 ms RTT, this can even become a problem for Cubic (see \pref{fig:delay_diff}, right).

\pref{fig:delay_diff} offers a more detailed view on RTT- and timeslice-dependent throughput for BBRv3 and Cubic.
Generally speaking, we again observe that BBRv3 can recover faster when the sender \ac{VM} has a 1 ms timeslice, which implies shorter off-CPU times and higher on-CPU frequency.
Longer RTT's are also beneficial, since the number of timeslices per 1 RTT increases.
Unexpectedly, when considering the 10 ms RTT case, the throughput for CPU shares of at least 60\% recovers faster with 20 ms timeslices, than with 10 ms timeslices.
For lower CPU shares, \ie during strictly CPU-limited throughput conditions, the configured \ac{RTT} has no notable effect on the achieved throughput, similar to what we observed for the bottleneck bandwidth parameter.

\textbf{Takeaway:}
The throughput of all BBR versions is highly inhibited under \ac{SCHEDULING} conditions, whilst Cubic remains robust. 
Limited BBR connections either never approach the available bandwidth because they immediately exit the Startup phase, or they initially send close to the limit but quickly downward-spiral when the delivery rate fails to keep up with the estimated bandwidth.
Strictly CPU-limited sending occurs at low CPU shares, primarily depends on the timeslice length, and is characterized by the fact that varying bottleneck bandwidths or path \acp{RTT} do not affect the achievable throughput. Instead, BBR throughput is strictly capped at values below 20 Mbps, more often below 10 Mbps, in the measured cases.

\section[Solution Approach]{Solution Approach \hspace{0.25em} \texorpdfstring{\IIISymbol}{}}
\label{sec:solution}

Motivated by our findings, we (1) define an indicator capable of detecting CPU-limited performance, and (2) patch BBR so that it can better utilize available timeslices during such conditions.

\subsection{Detecting CPU-limited Senders based on Inflight Deficit}
A useful detection approach should be simple and generalizable, and capable to reliably differentiate the CPU-limited performance from other throughput inhibitors, e.g., actual low bottleneck bandwidth, increased packet loss, or app-limited traffic.

\begin{figure}[t]
    \centering
    \includegraphics[width=\linewidth]{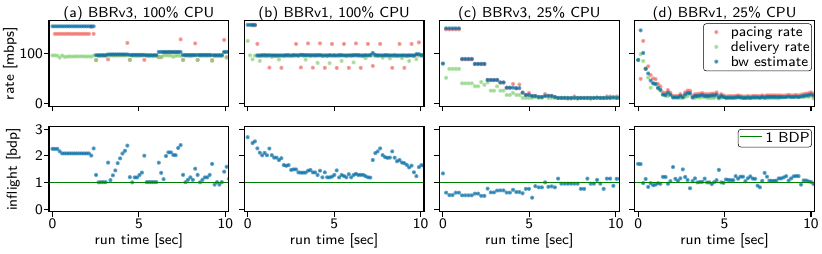}
    \caption{Correlation of BBRv1 and v3 data rates and inflight volume (in BDP as seen by BBR). This plot uses the same flow samples as \pref{fig:zoom3}, but adds the inflight metric.}
    \label{fig:rates_inflight}
\end{figure}

As elaborated in \pref{sec:background}, a TCP sender can only utilize the full bandwidth by having 1 \ac{BDP} of bytes inflight.
In case of BBR, the amount of data inflight is controlled by the pacing rate which sends packets at roughly the estimated bandwidth rate, \ie creating 1 BDP of bytes inflight. 
The upper limit per RTT is given by the \ac{cwnd} which is normally set to 2 \ac{BDP} to account for delayed ACKs~\cite{bbr_concept}.
Thus, \textbf{inflight data dropping below 1 BDP} should not happen during ProbeBW unless the sender is either app-limited (the network application does not generate enough data) or \textbf{CPU-limited}.
Thus, our detection approach relies on tracking the BDP-relative amount of data inflight, either in-application or via Linux utilities such as the \texttt{ss}-tool. 
When bytes inflight drop below BBR's \textit{currently estimated} \ac{BDP}, the socket is considered CPU-limited, as long as it is not app-limited.

As shown in \pref{fig:rates_inflight}, this "Inflight Deficit" approach is supported by our data set: 
While there are almost always 1-2 \ac{BDP} of bytes inflight in the case without \ac{SCHEDULING}, bytes inflight drops below the critical 1 \ac{BDP} threshold when the VM receives a 25\% CPU share, even after BBR has heavily decreased the bandwidth.
As elaborated above in \pref{sub:eval_overview}, this is because the sum of non-sending off-CPU time and non-sending inter-packet pacing offsets does not leave enough time to send at the expected data rate.
Note that the inflight volume is measured in \textit{BBR-estimated \ac{BDP}}, and not via a-priori known link characteristics.

\subsection{Inflight-based BBR Patch: Approach and Evaluation}
\label{sub:patch}
Having defined an indicator for CPU-limited throughput that can be determined by BBR based on \ac{BDP} and bytes-in-flight we now propose a patch\footnote{Code available under: \url{https://github.com/sys-uos/bbr-cpu-contention}} for the TCP BBR algorithm(s).

\paragraph{Approach}
As described in \pref{sec:background}, BBR data rates are driven by cwnd and pacing rate.
In contrast to the fix proposed by \cite{ha2021}, we do not modify the cwnd sizing since the cwnd will be automatically increased via the current \ac{BDP}, if the approach succeeds at maintaining a high bandwidth estimate. 
Further increasing the cwnd, \eg via cwnd\_gain, would lead to inflated cwnd, potentially compromising the fairness of BBR (cf.~\cite{zeynali2024}), especially seeing that prior studies identified BBR becoming cwnd-limited when competing with other flows~\cite{bbr_ware_2019}.
Thus, we aim at better utilizing the available CPU share by increasing the pacing rate via pacing\_gain.
To re-use established BBR variable settings, and keep the algorithm modification as minimal as possible, we choose BBR's "high gain" level, \ie 2.77 (BBRv1: 2.89), as pacing\_gain during CPU-limited periods.

To implement this approach, we modify the \texttt{bbr\_update\_gains()} function, as outlined by the pseudocode below, which only considers its simplified manipulation of pacing\_gain:

\begin{lstlisting}[language=C++, basicstyle=\scriptsize\ttfamily, mathescape]
bbr_update_gains(socket):
    pacing_gain = current_bbr_phase->gain;
    
    if socket->is_app_limited or current_bbr_phase $\neq$ ProbeBW: return;
    inflight = socket->bytes_sent - socket->bytes_acked;
    if inflight $< \alpha$ * current_bdp and (now-probe_rtt_done) $\geq$ $r$RTT:   // detect inflight deficit
            pacing_gain = HIGH_GAIN;  // overwrite pacing gain
\end{lstlisting}

We have stated above that a sub-BDP inflight volume could also be due to an app-limited sender.
Since we do not want to trigger the patch in this case - it could potentially introduce unfairness - we explicitly exclude app-limited sockets from our modification.
Other (\emph{false positive}) situations where ProbeBW could send below 1 BDP without CPU contention are right after ProbeRTT ended where inflight still recovers from the small cwnd, and -- for small bottleneck buffers -- in the DOWN cycle where BBR tries to drain queues.
The following parametrizations can be applied to the patch to prevent false positives but trade-off some performance gains:
Refrain from signaling an inflight deficit for up to $r$ RTT after ProbeRTT ended, and set the deficit threshold to $\alpha \leq$ 1 times BDP.

\begin{figure}
    \centering
        \includegraphics[width=\linewidth]{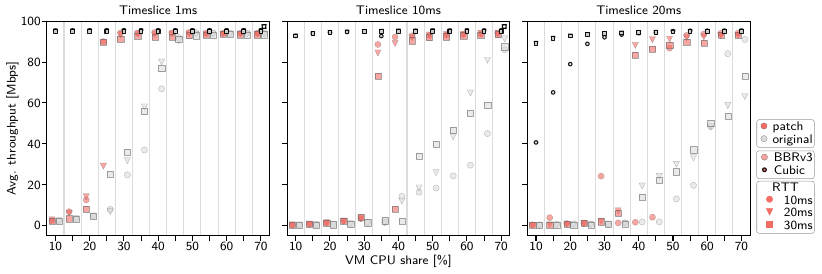}
    \caption{BBRv3 throughput medians comparing patch and original algorithm versions under \ac{SCHEDULING} conditions. Bottleneck bandwidth fixed at 100 Mbps.}
    \label{fig:eval_solution_agg}
\end{figure}

For BBRv3, we parametrize the patch so that diverse test runs do not produce false positives, by setting $r$ = 2 and $\alpha$ = 0.84 which is just below the inflight with headroom threshold in the CRUISE cycle \cite{bbr3}. 
Since BBRv1 uses a higher effective model update frequency, the trade-off between false positives and performance is more difficult to navigate. Due to space limitations, details on patch parametrization and evaluation can be found in the Appendix.

\paragraph{Validation without CPU contention}
To validate our patch and make sure that there are no false positives during non-CPU-limited periods, we start multiple $>$2-minute long iperf3 flows on the sender VM without emulating CPU contention. 
During this test a competing flow sharing the bottleneck link repeatedly starts and stops after a few seconds, and the RTT is varied between 10 and 30 ms during the connection.
Logging every time that BBR enters the patched lines, we can confirm that the patch was not activated during the monitored runtime.
Similarly, no false positives were recorded on a TCP socket that regularly becomes app-limited. 

\begin{figure}
    \centering
    \includegraphics[width=\linewidth]{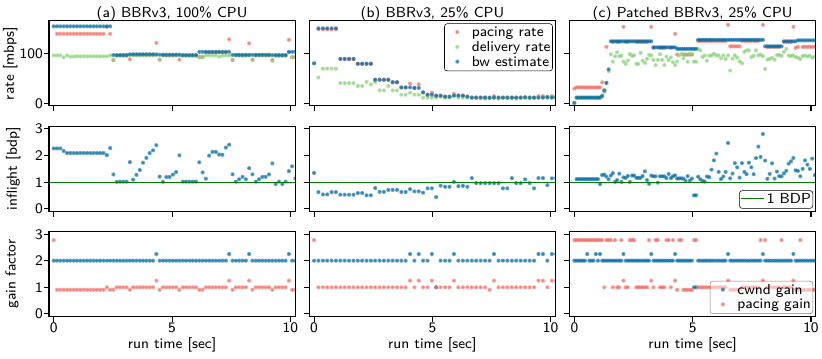}
    \caption{Correlation of BBRv3 data rates, inflight volume (in BDP as seen by BBR), and BBR's pacing\_gain and cwnd\_gain parameters. This plot uses the same flow samples as \pref{fig:zoom3} and \ref{fig:rates_inflight}, but adds the gain metric.}
    \label{fig:eval_solution_inflight}
\end{figure}

\paragraph{Performance Evaluation}
\pref{fig:eval_solution_agg} presents the results of patched BBRv3 versus the original implementation, at 100 Mbps bottleneck bandwidth for varying \acp{RTT} and timeslices. 
As shown, the patch is able to effectively mitigate the problem by shifting the problematic conditions to smaller CPU shares. In particular, patched BBRv3 can fully use 100 Mbps starting at 25\% CPU share for 1 ms timeslices, and at 40\% CPU share for 10 ms timeslices. 
In contrast, the original implementation needs at least 45\% and 70\% CPU shares, respectively, to achieve similarly high rates.
\pref{fig:eval_solution_inflight} shows that patched BBRv3 compensates inflight deficits (below the green line) by increasing the pacing gain, and returns to the normal pacing\_gain, as long as the inflight volume is at least 1 \ac{BDP}.
The \ac{RTT} measurements (not shown) are close to propagation delay for most of the connection which reflects the $\alpha$BDP threshold of the patch. 
Sudden and steep propagation delay variations in the path can affect the \ac{BDP}-based deficit detection: while smaller increases ($<$30 ms) do not impact the patch's performance during CPU contention, larger jumps can lead to temporal dips in throughput which are recovered after a few seconds. 

Reusing BBR's high gain constant and a threshold $\leq$ 1 BDP, it is not surprising that our patch can only mitigate the issue up to a certain point.
Performance during \ac{SCHEDULING} could be further improved by using higher pacing\_gain and a higher deficit threshold $\alpha$, but should be implemented with caution in regards to fairness.
Nevertheless, our evaluation indicates that increasing the pacing rate during inflight deficits is a promising approach to mitigate CPU-limited BBR throughput. 
In \pref{sec:real_kvm}, we will show that this insight still holds under real hypervisor scheduling conditions.

\paragraph{Fairness Evaluation}
While a thorough evaluation of fairness between CPU-limited (and non CPU-limited) flows is out of scope for this paper, we do want to make sure that our patch does not have unintended effects on fairness. 
For that, we run patched and original BBRv3 flows on two separate VMs sharing a 100 Mbps bottleneck link.
In our tests without CPU contention, we do not record any false positives for the patched version (\cf validation above), so that patched BBRv3 behaves just like BBRv3 (see \pref{fig:eval_fairness}).
We then subject one or both sender \acp{VM} to CPU contention (25\% - 50\% CPU share with 1 ms on-CPU timeslice).
When only the patched BBRv3 sender is CPU-limited, it is able to use close to 50\% of the available bandwidth in the 50\% CPU share setting (25\% CPU share setting: approx. 30\%), but does not grab an unfair share.
When both senders are CPU-limited, the patched version has an advantage over unpatched BBRv3 (\pref{fig:eval_fairness_hetero}), but the patch still cannot reach an equitable share when competing against Cubic (not shown) which uses 60\%-75\% of the bandwidth.
This unfairness is due to the varying CPU contention robustness of Cubic, patched BBRv3, and BBRv3, respectively.

\section{Real-World Validation on a KVM Hypervisor}
\label{sec:real_kvm}
\begin{figure}[t]
    \centering
    \includegraphics[width=\linewidth]{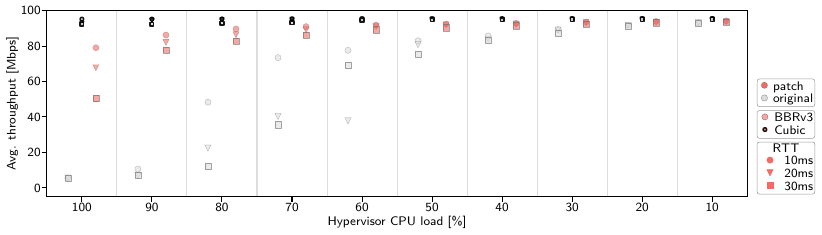}
    \caption{BBRv3 throughput medians comparing patch and original algorithm versions under real KVM scheduling conditions. Bottleneck bandwidth fixed at 100 Mbps.}
    \label{fig:kvm_results1}
\end{figure}

As mentioned earlier (\pref{sec:method}), our deadline-scheduling approach is intended to provide a controllable, repeatable, and hypervisor-independent emulation of CPU contention conditions.
The results for a given CPU share essentially yield the worst-case impact of CPU impact, due to the inherent inflexibility of deadline scheduling.
However, our method cannot reflect the complexities of real-world hypervisor scheduling or CPU usage patterns.
Thus, we now validate the emulated results under the real scheduling patterns of a KVM hypervisor during CPU contention.
Based on the TCP measurement setup described in \pref{sub:tcp_setup}, we start the sender \ac{VM} and a competing \ac{VM} on our KVM hypervisor. 
We artificially create CPU contention on the host by limiting the physical CPU resources of the hypervisor to a single physical core, and by generating x\% CPU load on the two \acp{VM} respectively.
\pref{fig:kvm_results1} shows BBRv3 performance under varying \acp{RTT} and artificial CPU load. In consistency with the emulated results, BBRv3 performance at lower CPU shares (= higher artificial CPU load) is heavily and increasingly degraded compared to Cubic under the same conditions.
Note that the x-axis shows the input parameter CPU load, since the VM's CPU share is a byproduct instead of a controllable variable. 
At 100\% CPU load, BBRv3 is limited to less than 10 Mbps, while patched BBRv3 reaches up to 80 Mbps for lower RTTs.
We have repeated the measurements for a higher bandwidth scenario (500 Mbps) as well (\cf Appendix \pref{fig:real_kvm_500}). As before, the most severe cases of BBRv3 degradation are bandwidth-independent but the recovery phase of the CPU-limited senders stretches to lower CPU load percentages.
The clear throughput improvement of the patch compared to the original algorithm remains, but again the recovery point is shifted to lower CPU load percentages.

In contrast to the emulated CPU contention framework which takes fine-granular scheduling parameters as input (see \pref{fig:method}), scheduling now becomes an output variable implicitly caused by the CPU load generation.
Due to this lack of direct control over the scheduling decisions, the "Real-KVM" measurements themselves are less generalizable and not ideal for characterizing the BBR performance degradation.
However, they highlight the real-world relevance of our emulated results. \newline

\section{Limitations}
\label{sec:limitations}
As described in \pref{sec:method}, our evaluation architecture emulates both the network and the CPU contention conditions.
Thus, it opens up a wide parameter space, especially considering that numerical parameters can be evaluated at very high resolutions and over a wide range.

In terms of network conditions, the evaluation presented in this paper cannot and does not intent to resemble a complex, realistic Internet setting e.g. with cross traffic lossy wireless links, short flows, or request-response workloads, but a controlled setup focusing on the most relevant network parameters related to BBR and CPU contention, in particular the \ac{RTT} and bottleneck bandwidth.
We confirmed that modest packet loss and differing bottleneck buffer sizes do not significantly impact CPU-limited BBR throughput via small scale measurements.

Regarding the CPU contention emulation, our choices for timeslice lengths and CPU shares are realistic in that they are based on known hypervisor behavior\footnote{Xen uses a default timeslice length of 30 ms~\cite{ha2021}, KVM has a default granularity of 2.8 ms in Ubuntu 24.04.} and VM products\footnote{At the time of writing, some VM products guarantee CPU shares of 5-10\%: \url{aws.amazon.com/ec2/instance-types/t4}}.
Still, by using a rather limited set of timeslice lengths for the evaluation (1,10,20ms), we might miss additional effects or dependencies. 

\section{Conclusion and Future Work}
Based on initial findings of BBR throughput degradation during \ac{SCHEDULING} in prior work, this paper provides a methodology to measure TCP throughput under controllable, repeatable \ac{SCHEDULING} conditions by "misusing" Linux deadline scheduling to fully control timeslice length and CPU share of the sender \ac{VM}.
This framework enabled our fine-granular evaluation, showing that BBRv1-v3 throughput is highly inhibited during \ac{SCHEDULING} under various network conditions, whilst Cubic remains robust. 
We identify a \textit{CPU-limited} socket state, where BBR throughput is capped below 10-20 Mbps, independently of the available link bandwidth or \ac{RTT}.
This state occurs up to a timeslice length-dependent CPU share threshold where BBR throughput starts to recover, until it can fully utilize the available bandwidth.
BBR throughput was only fully recovered at 45\% and 70\% CPU shares, for 1 ms and 10-20 ms timeslices, respectively.
Evaluating BBR state variables during a connection, we saw that CPU-limited BBR senders struggle to maintain 1 BDP inflight due to the pacing mechanism, which slows down the transmission during on-CPU times. This leads to a severe underutilization of the available bandwidth.

To mitigate this problem, we propose a BBR patch which monitors the socket's inflight data and (conservatively) increases the pacing rate when CPU-limited sending is detected. Our evaluation shows that this approach, whilst not solving the most severe cases of \ac{SCHEDULING}, can solve the problem in most scenarios by essentially shifting the critical threshold to lower CPU shares.

An interesting avenue for future work is to evaluate BBR throughput under CPU contention beyond Linux TCP, \ie using QUIC. Our preliminary measurements using several QUIC implementations suggest that differing pacing implementations and userspace scheduling add more variance to the problem.
Whilst our threshold based Inflight Deficit indicator allows detecting the problem from within VM guests, a challenge for future work is to detect CPU-limited BBR senders on the real Internet which requires correlating low throughput with CPU contention - difficult to do without access to the Cloud-based hosts systems.
Overall, due to today's high prevalence of both BBR and Cloud-based hosting, further research should be conducted on this topic to ensure the robustness of a BBR-based Internet.

\section*{Ethical Statement}
This work does not raise any ethical issues.

\newpage

\bibliographystyle{ACM-Reference-Format}
\bibliography{sample-base}

\appendix
\section*{Appendix}
\label{sec:appendix}

\begin{figure}[h]
    \centering
        \includegraphics[width=0.46\linewidth]{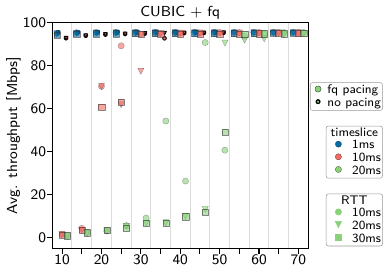}
    \caption{Paced (fq) vs. unpaced Cubic throughput (medians), for varying \ac{SCHEDULING} conditions and \acp{RTT}. Bottleneck bandwidth is fixed to 100 Mbps.}
    \label{fig:eval_cubic_fq}
\end{figure}

\begin{figure}[h]
    \centering
    \includegraphics[width=\linewidth]{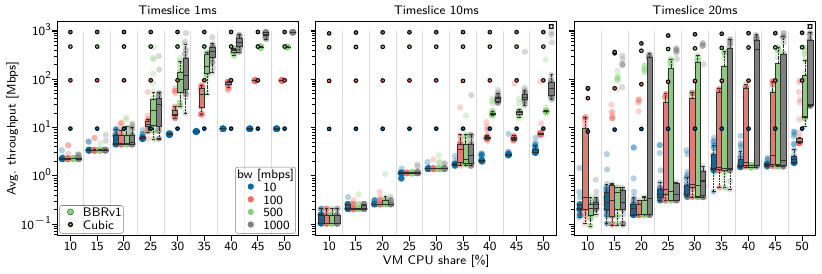}
    \caption{BBRv1 throughput at 10 ms RTT for varying bottleneck bandwidths. Note the logarithmic y-scaling.}
    \label{fig:rate_diff_bbrv1}
\end{figure}

\begin{figure}[h]
    \centering
    \includegraphics[width=\linewidth]{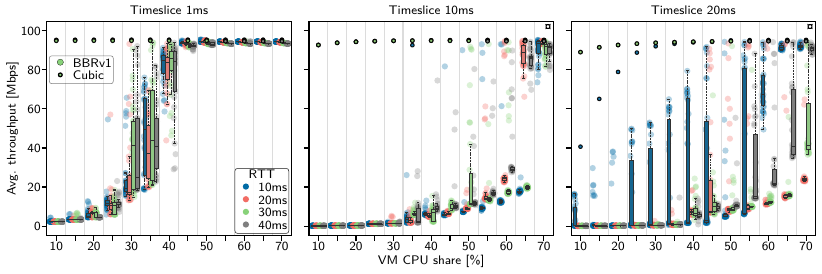}
    \caption{BBRv1 throughput at 100 Mbps bottleneck bandwidth for varying \acp{RTT}.}
    \label{fig:delay_diff_bbrv1}
\end{figure}

\begin{figure}[h]
    \centering
    \includegraphics[width=\linewidth]{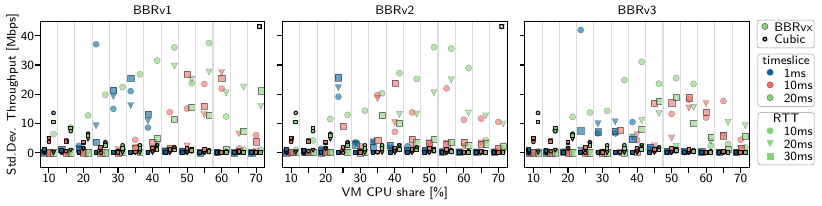}
    \caption{Standard deviation of per-run averaged throughputs for varying \ac{SCHEDULING} conditions and \acp{RTT}. Bottleneck
bandwidth is fixed to 100 Mbps.}
    \label{fig:overview_stddev}
\end{figure}

\subsection*{BBRv1 patch parametrization and results}
\begin{figure}[h]
    \centering
        \includegraphics[width=0.49\linewidth]{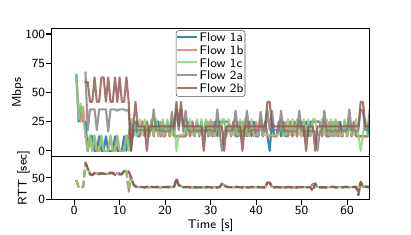}
        \includegraphics[width=0.49\linewidth]{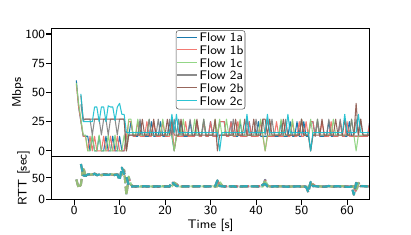}
    \caption{6 competing patched-BBRv1 flows on 2 separate sender \acp{VM} without (left), and without (right) CPU contention. Bottleneck bandwidth 100 Mbps, flows 2x start 3 seconds after flows 1x. Showing two representative samples.}
    \label{fig:eval_fairness_bbr1}
\end{figure}

When parametrizing the BBRv1 patch, as outlined in \pref{sub:patch}, we need to consider that this version reacts more quickly to a lower delivery rate in most situations, since it keeps a windowed maximum over 10 rounds ($\approx$ 10 \ac{RTT}), versus BBRv3 where the maximum remains for one ProbeBW cycle -- which is longer than 10 \ac{RTT} for most Internet-realistic \acp{RTT}.
This is visible in \pref{fig:rates_inflight} where BBRv3's performance goes down in steps, whereas BBRv1's throughput seems to degrade more continuously, updating the bandwidth estimate in smaller time intervals.
More frequent model updates imply more frequent \ac{BDP} updates so that our \ac{BDP}-dependent inflight deficit will never be large.
To prevent this, a BBRv1 patch must intervene faster, i.e. at a higher threshold $\alpha$, to prevent the degradation. Our experiments show that, especially for shorter time slices, only $\alpha$=1, can significantly improve performance during CPU-limited times.
However, this setting makes the algorithm more prone to false positives, specifically during cycles with pacing gain = 0.75 with small bottleneck buffers.
Since false positives cannot be ruled out, we need to pay attention to the fairness of this patch version. \pref{fig:eval_fairness_bbr1} shows that after a few seconds, 6 bottleneck-sharing flows achieve equitable throughputs with and without CPU contention. If contention occurs for some flows only, these CPU-limited flows are always disadvantaged towards the non CPU-limited flows, indicating that the activation of the patch allows for better performance but does not produce unfair behavior towards the original BBRv1 behavior.

\begin{figure}[h]
    \centering
        \includegraphics[width=\linewidth]{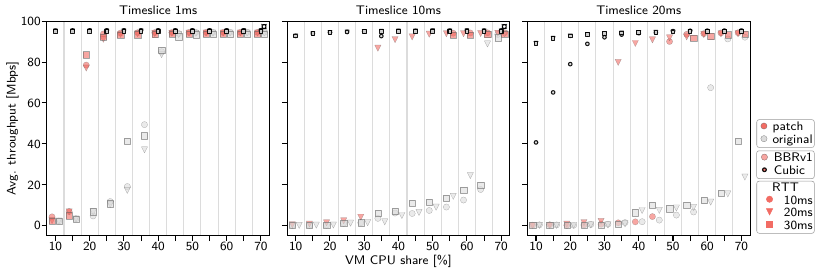}
    \caption{BBRv1 throughput medians comparing patch and original algorithm versions under \ac{SCHEDULING} conditions. Bottleneck bandwidth fixed at 100 Mbps.}
    \label{fig:eval_solution_agg_bbrv1}
\end{figure}

\begin{figure}[h]
    \centering
        \includegraphics[width=0.49\linewidth]{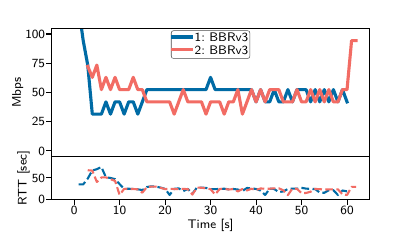}
        \includegraphics[width=0.49\linewidth]{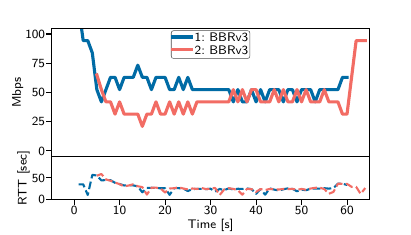}
    \caption{Two competing patched-BBRv3 flows on separate sender \acp{VM} without CPU contention. Bottleneck bandwidth 100 Mbps, flow 2 starts 3 seconds after flow 1. The patch mitigation was never not activated in either of the flows, so the fairness can be considered equivalent to the original BBRv3. Showing two representative samples. Note that prior work~\cite{zeynali2024} has shown that two staggered BBRv3 flows often take long to achieve equitable bandwidth sharing.}
    \label{fig:eval_fairness}
\end{figure}

\begin{figure}[h]
    \centering
        \includegraphics[width=0.49\linewidth]{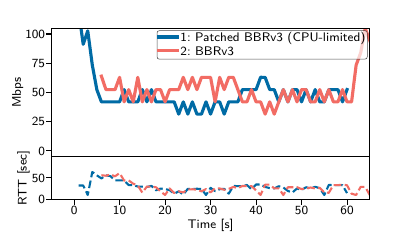}
        \includegraphics[width=0.49\linewidth]{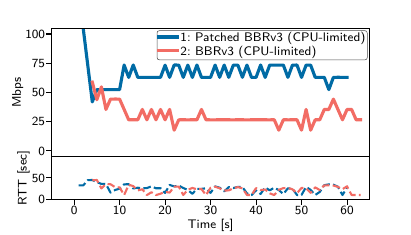}
    \caption{Two competing flows on separate sender \acp{VM} with one or both flows being slightly CPU-limited (50\% CPU share, 1 ms time slice). Bottleneck bandwidth 100 Mbps, flow 2 starts 3 seconds after flow 1. Showing representative samples. Note that prior work~\cite{zeynali2024} has shown that two staggered BBRv3 flows often take long to achieve equitable bandwidth sharing.}
    \label{fig:eval_fairness_hetero}
\end{figure}

\begin{figure}
    \centering
    \includegraphics[width=\linewidth]{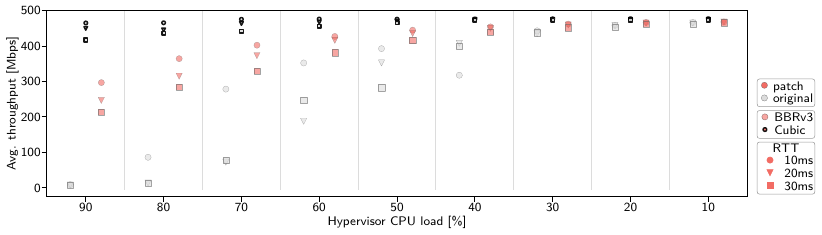}
    \caption{BBRv3 throughput medians comparing patch and original algorithm versions under real KVM scheduling conditions. Bottleneck bandwidth fixed at 500 Mbps.}
    \label{fig:real_kvm_500}
\end{figure}

\end{document}